\documentclass[12pt]{article}

\usepackage{latexsym}

\usepackage[font=small,labelfont=bf]{caption}

\usepackage{graphics}
\usepackage{graphicx}
\usepackage{epstopdf}
\usepackage{amssymb}
\usepackage{tabularx}
\usepackage{caption}
\usepackage{subcaption}
\usepackage{dcolumn}
\usepackage{bm}
\usepackage{hyperref}
\usepackage{tabularx}
\newcommand{\be}{\begin{equation}}
\newcommand{\ee}{\end{equation}}
\newcommand{\ba}{\begin{eqnarray}}
\newcommand{\ea}{\end{eqnarray}}
\newcommand{\ban}{\begin{eqnarray*}}
\newcommand{\ean}{\end{eqnarray*}}

\graphicspath{{./Figures/}}
\begin{document}


\title {On the shadow of rotating traversable wormholes}

\author{
Galin Gyulchev$^{1}$\footnote{E-mail: \texttt{gyulchev@phys.uni-sofia.bg}}, \, Petya Nedkova$^{1}$\footnote{E-mail: \texttt{pnedkova@phys.uni-sofia.bg}}, \, Vassil Tinchev$^{2}$,\\ Stoytcho Yazadjiev$^{1,2}$\footnote{E-mail: \texttt{yazad@phys.uni-sofia.bg}}\\ \\
   {\footnotesize${}^{1}$ Faculty of Physics, Sofia University,}\\
  {\footnotesize    5 James Bourchier Boulevard, Sofia~1164, Bulgaria }\\
  {\footnotesize${}^{2}$ Institute of Mathematics and Informatics,}\\
{\footnotesize Bulgarian Academy of Sciences, Acad. G. Bonchev 8, } \\
  {\footnotesize  Sofia 1113, Bulgaria}}
\date{}
\maketitle

\begin{abstract}
We revisit the shadow of rotating traversable wormholes discussing the role of the wormhole throat in the shadow formation. For certain classes of wormholes the throat serves as a potential barrier for light rays with particular impact parameters, thus modifying the shadow shape. We consider a couple of wormhole solutions and examine the structure of their shadow images, and the intrinsic mechanisms for their formation. Some of the shadows possess cuspy edges, which arise due to the interplay of two distinct families of unstable spherical orbits. These solutions provide examples, in which the explicit mechanism for cusp formation can be uncovered.
\end{abstract}

\section{Introduction}

An important phenomenological feature of black holes is their shadow, or their apparent shape when illuminated by a background source of light. Null geodesics in such spacetimes can be differentiated into two classes - trajectories which escape to infinity, and such that are captured by the compact object. Thus, a dark region appears in the observer's sky forming the shadow. The effect was originally discovered  when studying the light propagation in the vicinity of the Kerr black hole \cite{Bardeen}. Yet, it is not restricted exclusively to the presence of an event horizon, but can be observed in more general geometries possessing a photon region, such as naked singularities, gravastars and wormholes \cite{Nedkova:2013}-\cite{Cunha:2018}.

Motivated by present observational missions \cite{Doeleman}, the shadows of a variety of black holes were studied, both in general relativity and alternative theories of gravity \cite{Grenzebach:2014}-\cite{Cunha:2016a}. The specific features of the images can be used to extract information about the physical properties of the compact objects, and to differentiate between gravitational theories \cite{Johannsen}-\cite{Bambi:2015}. Using the deformation of the shadow shape of an isolated black hole we can evaluate its spin and multipole moments \cite{Tsukamoto}-\cite{Li}. On the other hand, we can detect the interaction of the black hole with an external gravitation field, since it leads to the formation of complicated lensing patterns with hierarchies of multiple shadow images \cite{Yumoto}-\cite{Grover:2018}.

The aim of the current work is to investigate the shadows of horizonless objects by studying the light propagation in a couple of rotating wormhole solutions. The wormhole spacetimes, which we consider, allow separation of variables for the null geodesics equations. Hence, we can obtain the shadow boundary analytically as a lensed image of the unstable spherical photon orbits. The unstable spherical photon orbits for our solutions divide into two distinct families - inner spherical orbits located at the wormhole throat, and outer ones, located at some larger radial distance. An analytical expression for the shadow boundary associated with the outer spherical orbits was obtained previously in \cite{Nedkova:2013}, but the shadow analysis presented there was incomplete. A second work \cite{Shaikh} derived the shadow rim resulting from the spherical orbits at the throat and presented the wormhole images as a superposition of the two boundary curves. Unfortunately, due to improper estimation of the wormhole mass, the shadow images reported in \cite{Shaikh} deviate from the realistic ones both qualitatively, and in terms of scale. Therefore, we revisit the problem by summarizing current results, and computing the wormhole shadows both analytically by means of the boundary curves, and numerically using a ray-tracing program. In addition, we perform a detailed analysis of the structure of the shadow images, and the internal mechanisms underlying their formation. Considerable intuition for the qualitative behavior of the shadow when varying the angular momentum is obtained by the investigation of the unstable light rings.

One of the wormhole solutions, which we study, is characterized by a cuspy structure of the shadow edge. Using the analytical expressions for the shadow boundary, we were able explain the formation of the cusps. They arise at the junction of the curves determined from the two families of spherical orbits due to a non-smooth merger of the two branches. While it was anticipated in \cite{Cunha:2017b}-\cite{Wang} that the cusps result from the interplay between stable and unstable spherical photon orbits, we provide for our solution the explicit mechanism for their formation. For an equatorial observer, we investigate the behavior of the cusps when varying the angular momentum proving that a critical spin parameter exists, for which they disappear. It coincides with the regime when one of the families of unstable spherical orbits ceases to give contribution to the shadow boundary.

The paper is organized as follows. In section 2 we review the class of rotating traversable wormhole solutions, which we study, and the null geodesic equations in the corresponding spacetimes. The analytical curves following from the spherical photon orbits, which determine the shadow boundary, are presented. In section 3
we study the shadows of two particular wormhole solutions by analytical and numerical means. In certain cases we observe cuspy shadow edges, the structure and dynamics of which we analyze in detail for various angular momenta. Section 4 is devoted to conclusion.

\section{Rotating traversable wormholes}

Rotating traversable wormholes can be described by a general stationary and axisymmetric metric, proposed by Teo in the form \cite{Teo:1998}

\begin{equation}\label{metric}
ds^2=-N^2{\rm d}t^2+\left(1-{b\over r}\right)^{-1}{\rm d}r^2+r^2K^2
\left[{\rm d}\theta^2+\sin^2\theta({\rm d}\varphi-\omega{\rm d}t)^2
\right],
\end{equation}
where the functions $N$, $b$, $\omega$ and $K$ depend only on the spherical coordinates $r$ and $\theta$. The metric functions should satisfy certain conditions, which ensure the regularity of the solution, and lead to the characteristic geometry and physical properties of the wormhole. For example, the redshift function $N$ should be finite and nonzero, the shape function $b$ should obey the restrictions $\partial_\theta b(r,\theta)=0$ and  $\partial_r b(r,\theta)<1$ at the wormhole throat, while the derivatives $\partial_\theta N(r,\theta)$, $\partial_\theta b(r,\theta)$ and $\partial_\theta K(r,\theta)$ should vanish on the rotation axis $\theta=0$ and $\theta = \pi$. In addition,  asymptotically flat wormholes should possess the following expansions at the physical infinity $r\rightarrow \infty$

\begin{eqnarray}\label{asympt}
&&N = 1 - \frac{M}{r} + {\rm O}\left({1\over r^2}\right), \quad ~~~ K = 1 + {\rm O}\left({1\over r}\right) , \quad~~~ \frac{b}{r} = {\rm O}\left({1\over r}\right), \nonumber \\
&&\omega={2J\over r^3}+{\rm O}\left({1\over r^4}\right),
\end{eqnarray}
which involves the conserved charges corresponding to the physical characteristics of the solution - the ADM mass $M$ of the wormhole, and its angular momentum $J$.

Despite these restrictions, the rotating wormhole geometry is general enough to include various particular cases with different physical properties.  We can distinguish an interesting class of solutions by requiring that all the metric functions $N$, $b$, $\omega$ and $K$ depend only on the radial coordinate. Then, we can separate the variables in the geodesic equations, and investigate the properties of the light propagation in the wormhole geometry by analytical means \cite{Nedkova:2013}.

Stationary and axisymmetric spacetimes admit two conserved quantities for geodesic motion,  associated with the particle's energy $E$ and angular momentum $L$. For separable geodesic equations an additional integral of motion arises, which we denote by $Q$. Then, introducing the impact parameters $\xi = L/E$, and $\eta = Q/E^2$,  we can write the null geodesic equations in the form \cite{Nedkova:2013}

\begin{eqnarray}\label{geodesic}
&&\frac{N}{\sqrt{1-\frac{b}{r}}}\frac{dr}{d\lambda} = \sqrt{R(r)}, \quad~~~
r^2K^2\frac{d\theta}{d\lambda} = \sqrt{T(\theta)}, \nonumber \\[2mm]
&&N^2\frac{d\varphi}{d\lambda} = \omega(1 - \omega\xi) + \frac{N^2~\xi}{r^2K^2\sin^2\theta}, \nonumber \\[2mm]
&&N^2\frac{dt}{d\lambda} = 1 - \omega \xi,
\end{eqnarray}
where we define the functions
\begin{eqnarray}
R(r) &=& \left(1-\omega \xi\right)^2 - \eta \frac{N^2}{r^2K^2}, \nonumber \\
T(\theta) &=& \eta - \frac{\xi^2}{\sin^2\theta}.
\end{eqnarray}

The radial geodesic equation can be interpreted effectively as one-dimensional motion in the field of the potential $V_{eff}$
\begin{eqnarray}
\left(\frac{dr}{d\lambda}\right)^2 + V_{eff} = 1  , \quad V_{eff} = 1 - \frac{1}{N^2}\left(1-\frac{b}{r}\right)R(r).
\end{eqnarray}
According to their impact parameters the null geodesics are separated into two classes - photon trajectories, which pass through the wormhole, and trajectories, which scatter away and escape to infinity. The boundary between the two classes is determined by a family of unstable spherical orbits satisfying the conditions

\begin{eqnarray}
 V_{eff} = 1,\quad~~~\frac{V_{eff}}{dr} = 0,\quad~~~\frac{d^2 V_{eff}}{dr^2} \leq 0.
\end{eqnarray}
They lead to a relation between the impact parameters $\xi$ and $\eta$, which defines a curve in the parameter space corresponding to the shadow rim. Introducing the celestial coordinates

\begin{eqnarray}\label{alpha}
\alpha &=& \lim_{r\rightarrow \infty}\left( -r^{2}\sin\theta_{0}\frac{d\varphi}{dr}\right)=-\frac{\xi}{\sin\theta_0}, \nonumber \\[2mm]
\beta &=& \lim_{r\rightarrow \infty}r^{2}\frac{d\theta}{dr} =\left(\eta - \frac{\xi^2}{\sin^2\theta_0}\right)^{1/2},
\end{eqnarray}
we obtain its projection on the observers's sky, i.e. the apparent position of the shadow boundary for an observer located at asymptotic infinity with an inclination angle $\theta_0$\footnote {The inclination angle is the angle between the rotation axis of the wormhole and the line of sight of the observer.}.

The unstable spherical orbits, which determine the shadow boundary, can be divides into two types. The first family results from a maximum of the effective potential located outside the wormhole throat. In this case the metric function $g^{rr} = (1-\frac{b}{r})$ is distinct from zero, and the spherical orbits satisfy the conditions
 \begin{eqnarray}
R(r) = 0,\quad~~~ \frac{dR}{dr}=0,\quad~~~\frac{d^2R}{dr^2}\geq0. \nonumber
\end{eqnarray}

These equations lead to the following parametric relations for the impact parameters derived in \cite{Nedkova:2013}

\begin{eqnarray}\label{curve1}
&&\eta = \frac{r^2K^2}{N^2}(1-\omega\xi)^2, \nonumber\\
&&\xi = \frac{\Sigma}{\Sigma\omega-\omega'}, \quad~~~ \Sigma = \frac{1}{2}\frac{d}{dr}\ln\left(\frac{N^2}{r^2K^2}\right).
\end{eqnarray}

The other family of spherical orbits results from a maximum of the effective potential located at the wormhole throat. In this case the metric function $g^{rr}$ vanishes, and the trajectories are determined by the conditions

\begin{eqnarray}
1-\frac{b(r)}{r}=0, \quad~~~ R(r)=0, \quad~~~ \frac{dR}{dr}\geq 0. \nonumber
\end{eqnarray}

They lead to the implicit relation between the celestial coordinates $\alpha$ and $\beta$ \cite{Shaikh}

\begin{equation}\label{curve2}
(\omega^2r^2_0 K^2\sin^2{\theta_0} - N^2)\alpha^2 + 2\omega r^2_0 K^2 \sin{\theta_0}\alpha + r^2_0 K^2 - N^2\beta^2\mid_{r_0} =0,
\end{equation}
where $r_0$ is the value of the radial coordinate at the wormhole throat.

In our analysis we consider a simple class of solutions with metric functions
\begin{eqnarray}\label{wormhole0}
b(r)=r_0, \quad~~~ K =1,\quad~~~ \omega = \frac{2J}{r^3},
\end{eqnarray}
while we allow for different choices of the redshift function $N(r)$, investigating its influence on the wormhole shadow. In the following discussion we use the spin parameter of the wormhole, defined as $a = J/M^2$, as a measure of the rotation rate.

\section{Wormhole shadows}


We consider two wormhole solutions with different redshift functions $N$ belonging to the class we described, which possess equal masses. We investigate the structure of the shadow in each case and its deformation for various angular momenta. The shadow images are constructed in two independent ways comparing the results. On the one hand the shadow boundary, which results from the analytical curves given by Eqs. ($\ref{curve1}$)-($\ref{curve2}$),  is illustrated. On the other hand the shadow is computed numerically by a ray-tracing program integrating the geodesic equations and separating  escape from plunge orbits. The analytic treatment gives the advantage that in each case we can get intuition about the intrinsic mechanism for the shadow formation. While certain differences can be extracted, the images in the two cases result from qualitatively similar structures.

\subsection*{Case I: $N = \exp{\left(-\frac{r_0}{r}\right)}$}

The first wormhole solution, which we consider, is characterized by a redshift function $N = \exp{\left(-\frac{r_0}{r}\right)}$. Expanding the redshift function near infinity, and comparing with the general asymptotic behavior for asymptotically flat solutions ($\ref{asympt}$),  we obtain that the wormhole mass $M$ is equal to the solution parameter $r_0$. The  shadow boundary is determined by the superposition of the two curves given by eqs. ($\ref{curve1}$)-($\ref{curve2}$). Each curve corresponds to a family of unstable spherical photon orbits, while their intersections with the equatorial plane determine a pair of planar circular orbits, also called in the literature light rings. The inner light ring is located at the wormhole throat, and outer one is at radial distance $r>r_0$.

\begin{figure}[htp]
    		\setlength{\tabcolsep}{ 0 pt }{\footnotesize\tt
		\begin{tabular}{ c }
            \includegraphics[width=\textwidth]{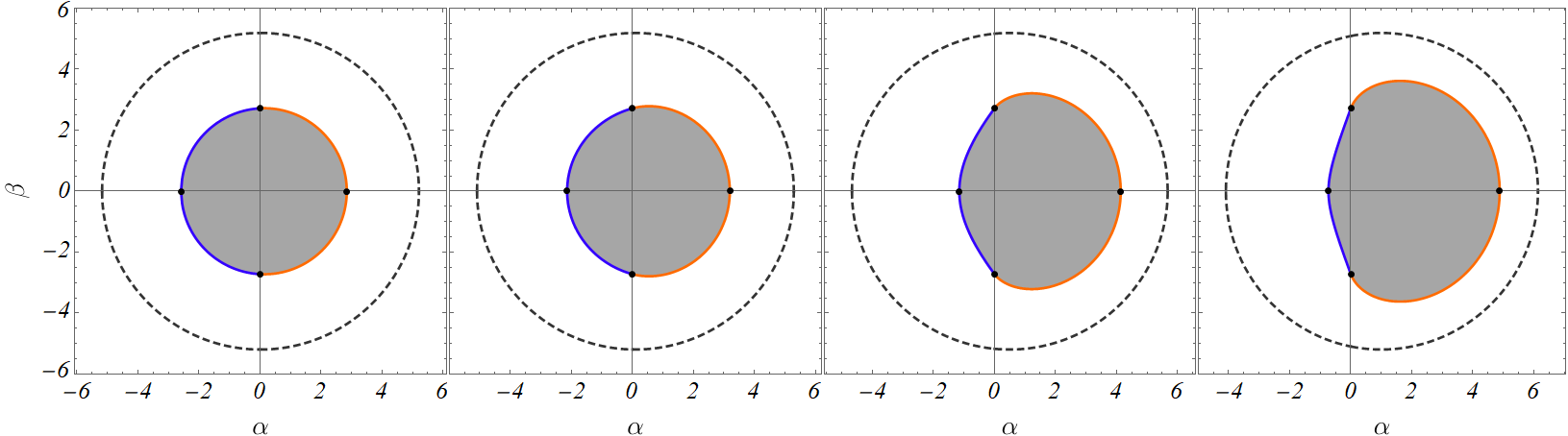} \\[1mm]
			\hspace{0.5cm} $a=0.01$ \hspace{2.1cm} $a=0.05$ \hspace{2.1cm} $a=0.25$ \hspace{2.1cm} $a=0.5$ \\[2mm]
            \includegraphics[width=\textwidth]{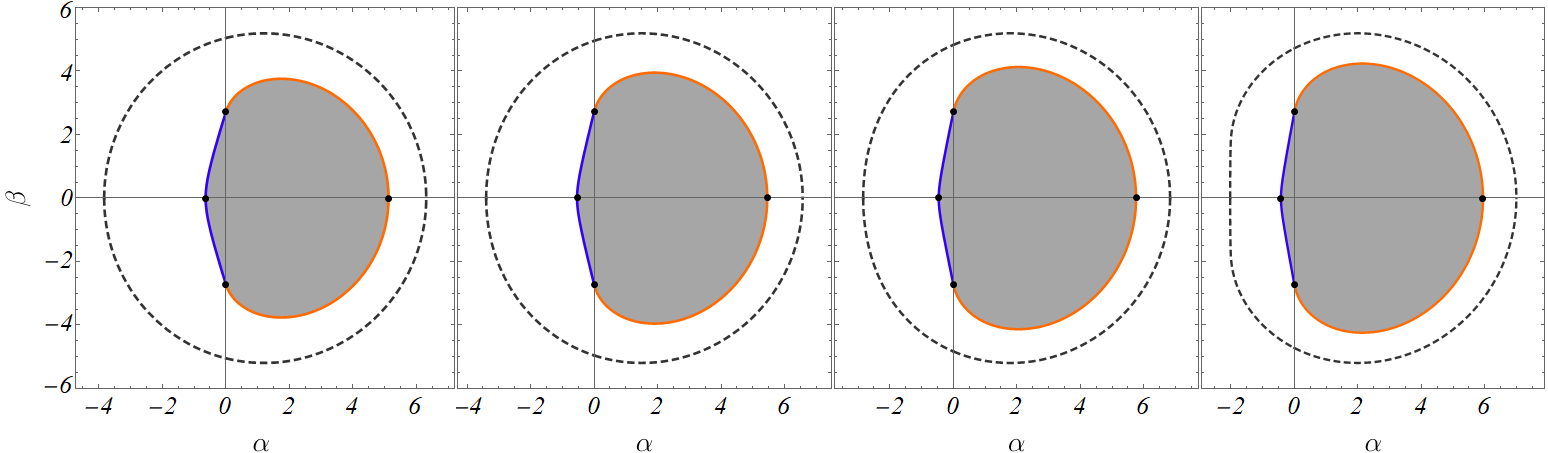} \\[1mm]
            \hspace{0.5cm} $a=0.6$ \hspace{2.25cm}  $a=0.75$ \hspace{2.25cm} $a=0.9$ \hspace{2.25cm} $a=1$
		\end{tabular}}
 \caption{\label{fig:S1_90}\small Wormhole shadow for the redshift function $N = \exp(-\frac{r_0}{r})$, and observer inclination angle $\theta_0 = 90^\circ$. The curves associated with the light ring at the throat, and the outer light ring are depicted in blue and orange, respectively. The dots denote the left- and rightmost edges of the shadow, evaluated analytically according to Eqs. ($\ref{aL_S1}$)-($\ref{aR_S1}$), and the intersection points of the two branches. The shadow boundary of the Kerr black hole is presented for comparison with black dashed line. The mass of the solutions is set equal to 1.}
\end{figure}

\begin{figure}[htp]
    		\setlength{\tabcolsep}{ 0 pt }{\footnotesize\tt
		\begin{tabular}{ c }
            \includegraphics[width=\textwidth]{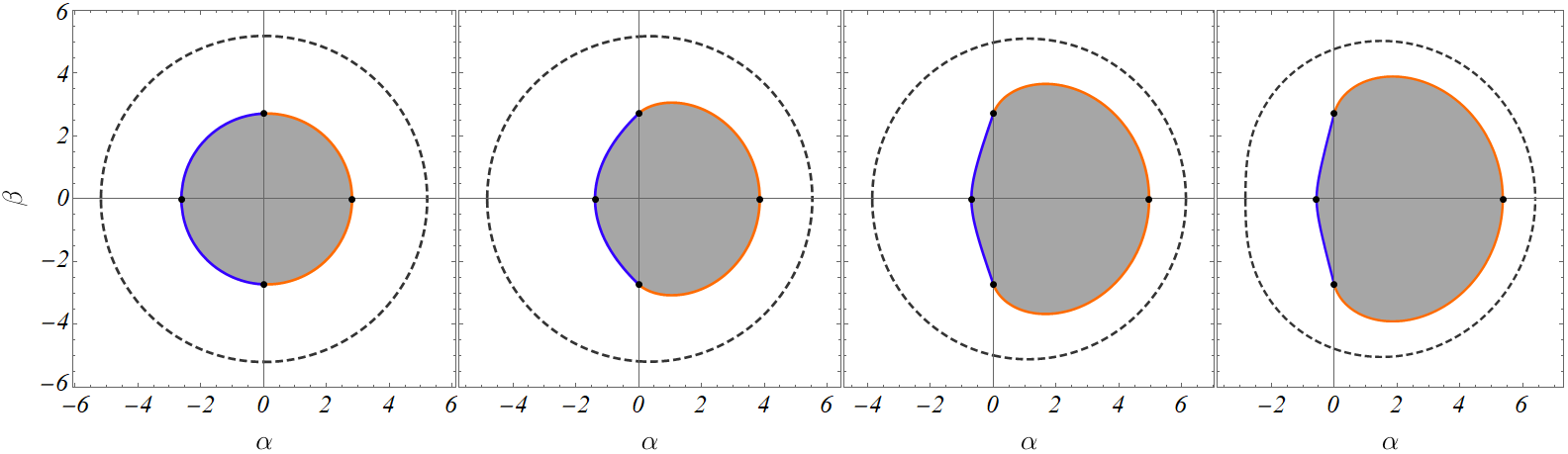}\\[1mm]
			\hspace{0.45cm} $a=0.01$ \hspace{2.15cm}  $a=0.25$ \hspace{2.15cm} $a=0.75$ \hspace{2.15cm} $a=1$
		\end{tabular}}
 \caption{\label{fig:S1_45}\small Wormhole shadow for the redshift function $N = \exp(-\frac{r_0}{r})$, and observer inclination angle $\theta_0 = 45^\circ$. The same conventions are used as in Fig. $\ref{fig:S1_90}$. The shadow boundary of the Kerr black hole is presented for comparison with black dashed line. }
\end{figure}

\begin{figure}[b!]
    \centering
		\setlength{\tabcolsep}{ 0 pt }{\footnotesize\tt
		\begin{tabular}{ c }
            \includegraphics[width=\textwidth]{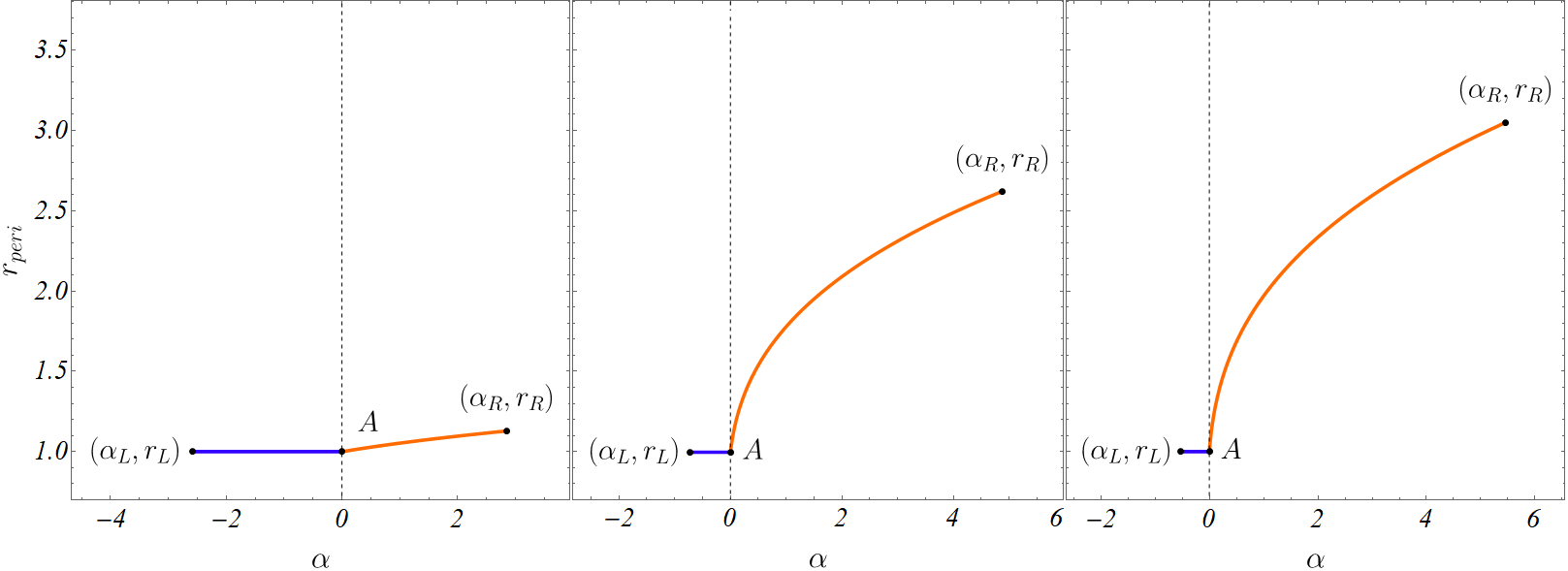}\\[1mm]
			\hspace{0.45cm} $a=0.01$ \hspace{3.35cm} $a=0.5$ \hspace{3.35cm} $a=0.75$
		\end{tabular}}
\caption{\label{fig:S1_FPO}\small Dependence of the impact parameter $\alpha$ on the radial distance along the shadow boundary for the redshift function $N = \exp(-\frac{r_0}{r})$. The impact parameters of the left- and rightmost shadow edges are denoted by $\alpha_L$ and $\alpha_R$, respectively, while the point $A$ represents the intersection point between the two shadow boundary branches. We denote by $r_{peri}$ the radial coordinate normalized by the location of the wormhole throat $r_0$. The observer inclination angle is $\theta_0 = 90^\circ$.}
\end{figure}

Examining the explicit form of the expressions ($\ref{curve1}$)-($\ref{curve2}$), we obtain that the curve resulting from the unstable light ring at the wormhole throat is defined only for negative values of the impact parameter $\alpha$, while the curve associated with the outer light ring exists only for positive values of $\alpha$. The shadow is presented in Figs. $\ref{fig:S1_90}$-$\ref{fig:S1_45}$, where the two branches forming the shadow boundary are depicted in blue and orange. The two branches merge at $\alpha = 0$ for the same value of the radial coordinate $r=r_0$, which can be seen in Fig. $\ref{fig:S1_FPO}$ illustrating the dependence of the impact parameter $\alpha$ on the radial distance along the shadow rim.

We can get an estimate of the shadow size by evaluating the intersection points of the shadow boundary with the coordinate axes $\beta =0$ and $\alpha=0$. The leftmost edge of the shadow is obtained by calculating the crossing point of the unstable branch of the curve ($\ref{curve2}$) with the equatorial plane, giving for the value of coordinate $\alpha$ the expression
\begin{equation}\label{aL_S1}
\alpha_L = -\frac{r_0}{2a\sin{\theta_0} + e^{-1}}.
\end{equation}
which depends on the spin parameter $a$ of the wormhole. Increasing the angular momentum, the value of $\alpha_L$ increases. Thus, the leftmost edge moves to the right, starting from the position of the photon sphere in the static case  $\alpha^{st}_L = -r_0e$, and approaching asymptotically $\alpha =0$ for large angular momenta.

In a similar way the rightmost edge of the shadow is calculated by the crossing point of the curve determined by the outer light ring ($\ref{curve1}$) with the axis $\beta =0$. The crossing point occurs for values of the parameter along the curve $r = r_R(a, \theta_0)$ corresponding to a solution of the algebraic equation

\begin{equation}\label{aR_S1}
\left(1-\frac{r_0}{r}\right)\frac{N}{\sin{\theta_0}} = 6a\frac{r^2_0}{r^2}.
\end{equation}
It can be shown that for any spin parameter and inclination angle the equation possesses a single root $r_R$ in the range $r_R > r_0$. Using ($\ref{curve1}$), we can calculate the value of the impact parameter $\alpha_R$ corresponding to it.

Another characteristic pair of points are the intersection points of the two curves, which form the shadow rim. Their coordinates are given by $(\alpha=0, \beta = \pm r_0 e)$, showing that they are independent of the spin parameter of the solution. Thus, increasing the angular momentum, the leftmost edge of the shadow approaches asymptotically $\alpha=0$, while the intersection points with the right branch stay  intact. Consequently, we expect that for large angular momenta the contribution to the shadow boundary of the curve connected with the light ring at the throat will become less and less significant.

The behavior of the shadow when varying the wormhole spin is illustrated in Figs. $\ref{fig:S1_90}$-$\ref{fig:S1_45}$ for two values of the observer's inclination angle.  They confirm the intuition gained from the investigation of the shadow edges. Due to the frame dragging, for high spins the curve determined by the light ring at the wormhole throat tends asymptotically to a straight line coinciding with the $\beta$-axis. The images are compared with the shadow of the Kerr black hole with the same mass. The wormhole shadows are in general smaller, but increasing the angular momentum the shadow area for positive $\alpha$ grows more rapidly than that of the Kerr black hole, so the deviation between the two images decreases. For extremal spin of the Kerr black hole the shadows of the two solutions are qualitatively very similar, although differing in size.

The influence of the wormhole throat on its shadow was investigated also previously in \cite{Shaikh}. However, due to improper estimation of the wormhole mass, the contribution of the light ring at the throat to the shadow boundary is overestimated, resulting in much rounder images. In addition, as a result of the same inaccuracy, the wormhole images reported in \cite{Shaikh} are bigger, approaching the shadow of the Kerr black hole in size.

\subsection*{Case II: $N = \exp{\left(-\frac{r_0}{r}-\frac{r^2_0}{r^2}\right)}$}

We consider another wormhole solution with the same mass $M =r_0$, characterized by a redshift function  $N = \exp{\left(-\frac{r_0}{r}-\frac{r^2_0}{r^2}\right)}$. In general, its shadow is determined again by two families of spherical orbits associated with the light ring at the throat, and the outer light ring. Contrary to the previous case, the two families intersect for negative values of the impact parameter $\alpha$. Thus, the spherical orbits arising from the outer light ring can occur both for positive and negative $\alpha$. The junction between the two families is not smooth, which leads to the formation of cusps in the shadow boundary. Below a certain critical angular momentum, which depends on the observer's inclination angle, the spherical orbits associated with the throat cease to give contribution to the wormhole shadow, and it is determined completely by the outer light ring. The qualitative behavior of the shadow image with respect to the variation of the wormhole spin can be separated in three distinct classes, which we describe below for the case of an equatorial observer.

\subsubsection*{1. Wormhole spin $a\geq a^{(1)}_{crit}\approx 0.07$}

\begin{figure}[t!]
    		\setlength{\tabcolsep}{ 0 pt }{\footnotesize\tt
		\begin{tabular}{ c }
           \includegraphics[width=\textwidth]{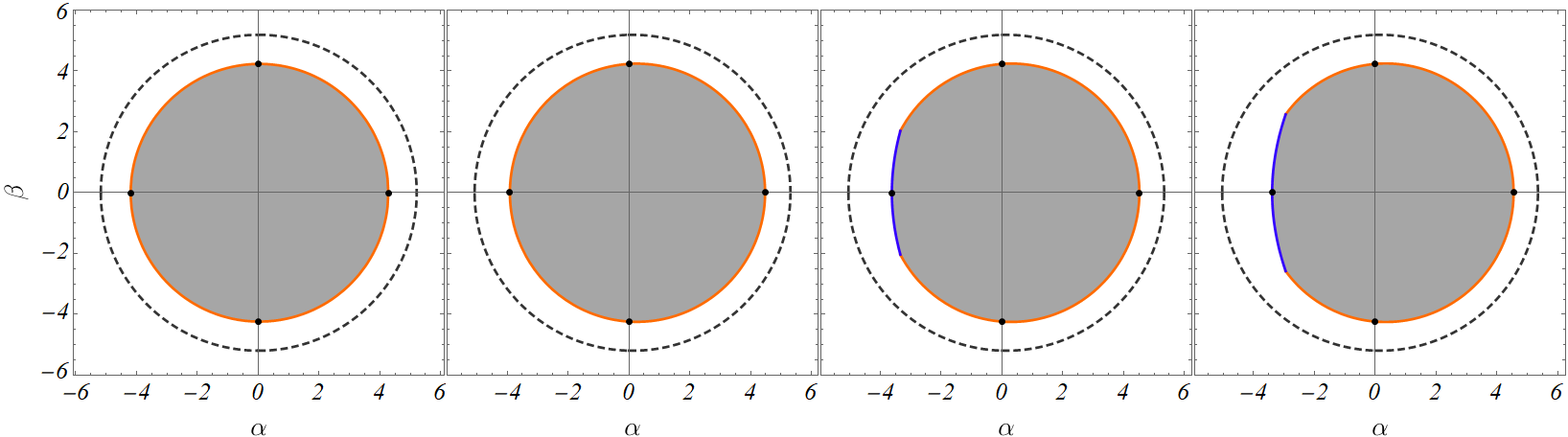} \\[1mm]
			\hspace{0.5cm} $a=0.01$ \hspace{2.1cm}  $a=0.06$ \hspace{2.1cm} $a=0.07$ \hspace{2.1cm} $a=0.08$ \\[2mm]
             \includegraphics[width=\textwidth]{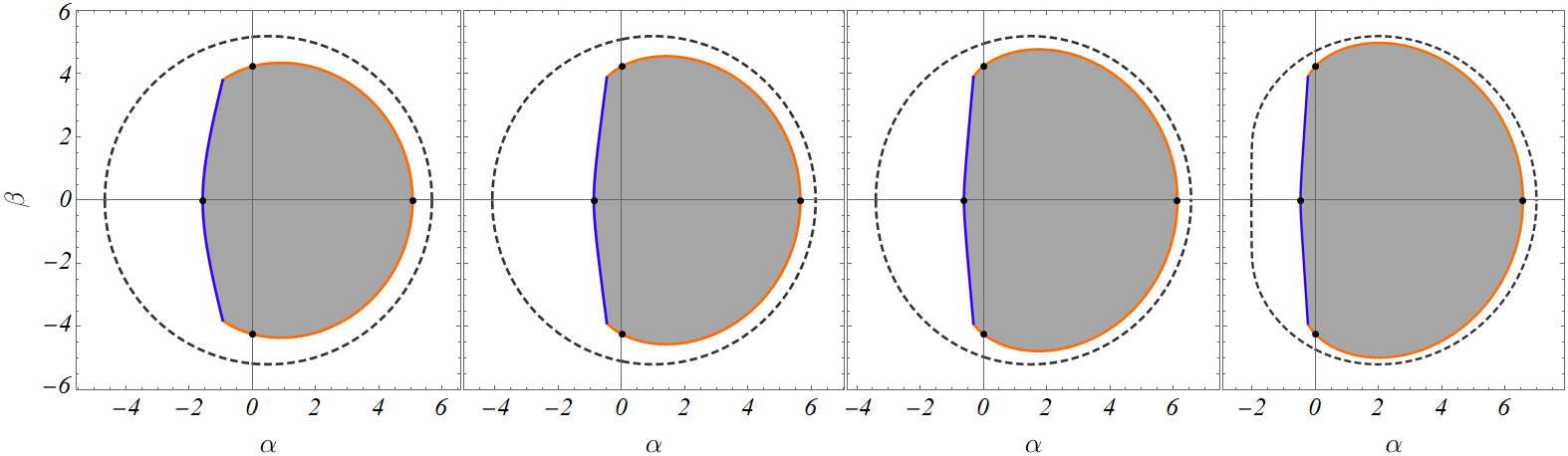} \\[1mm]
            \hspace{0.45cm} $a=0.25$ \hspace{2.18cm}  $a=0.5$ \hspace{2.225cm} $a=0.75$ \hspace{2.15cm} $a=1$
		\end{tabular}}
 \caption{\label{fig:S2_90}\small Wormhole shadow for the redshift function $N = \exp(-\frac{r_0}{r}-\frac{r^2_0}{r^2})$, and observer inclination angle $\theta_0 = 90^\circ$. The curves associated with the light ring at the throat, and the outer light ring are depicted in blue and orange, respectively. The dots denote the left- and rightmost edges of the shadow, evaluated analytically according to Eqs. ($\ref{aL_S2}$)-($\ref{aR_S2}$), and the intersection points with the axis $\alpha = 0$. The shadow boundary of the Kerr black hole is presented for comparison (black dashed line). The mass of the solutions is set equal to 1. }
\end{figure}

The shadow boundary is formed by the two branches ($\ref{curve1}$)-($\ref{curve2}$) determined by the two families of unstable spherical orbits (see Fig. ($\ref{fig:S2_90}$)). Its leftmost edge corresponds to the intersection of the curve associated with the spherical orbits at the throat with the equatorial plane given explicitly by

\begin{equation}\label{aL_S2}
\alpha_L = -\frac{r_0}{2a + e^{-2} }.
\end{equation}
In a similar way, the rightmost edge is located at the intersection of the curve representing the outer family of spherical orbits with the axis $\beta=0$. It corresponds to a radial distance solution to the equation

\begin{equation}\label{aR_S2}
\left(1-\frac{r_0}{r} -\frac{2r^2_0}{r^2}\right)N = 6a\frac{r^2_0}{r^2}.
\end{equation}

In the spin parameter range, which we consider, the equation possesses a single root $r>r_0$, which determines the impact parameter $\alpha_R$ at the right edge. The two branches intersect at negative values of $\alpha$ closing the shadow rim. Increasing the angular momentum, the values of $\alpha_L$ and $\alpha_R$ increase, and the two shadow edges move to the right. At the same time the intersection points between the two shadow branches approach the value $\alpha = 0$. As a result, for high spins the curve determined by the spherical orbits at the wormhole throat tends to a straight line coinciding with the $\beta$-axis.

Contrary to the previous wormhole solution with the redshift function $N = \exp{\left(-\frac{r_0}{r}\right)}$, the intersection point between the two shadow branches occurs for different values of the radial coordinate along the two curves. The whole left branch is located at $r=r_0$, which corresponds to the wormhole throat, while the radial distance on the right branch obeys $r_A > r_0$ at the intersection point. This behaviour leads to the formation of cusps in the shadow boundary. The phenomenon is previously observed for numerical solutions describing Kerr black hole with Proca hair \cite{Cunha:2017b}, and in the shadow of the black holes investigated in \cite{Wang}. In a similar way for these solutions the cusps arise due to the interplay between different spherical orbits.

\begin{figure}[t]
 \centering
		\setlength{\tabcolsep}{ 0 pt }{\normalsize\tt
		\begin{tabular}{ c }
            \includegraphics[width=0.8\textwidth]{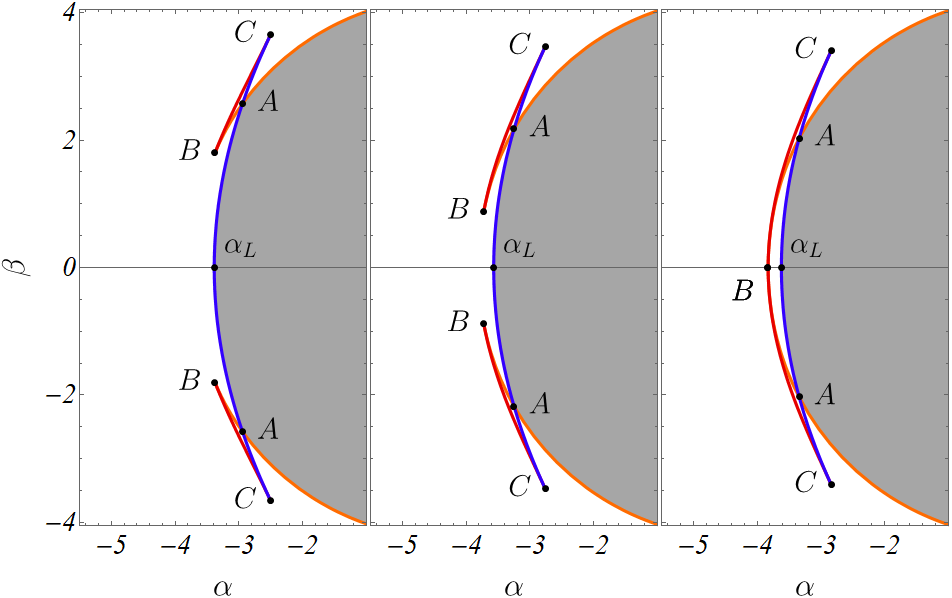} \\[2mm]
          		 \hspace{1.5cm}	${a=0.08}$ \hspace{1.8cm} $a=0.075$ \hspace{1.2cm} $a=a^{(1)}_{crit}\approx0.07$
		\end{tabular}}
\caption{\small Behavior of the spherical orbits in the vicinity of the shadow cusp for spin parameter $a\geq a^{(1)}_{crit}\approx 0.07$. The blue and orange lines represent the unstable spherical orbits associated with the two light rings, while their intersection point is denoted by $A$. The spherical orbits corresponding to the outer black ring become stable at the point $B$, and the stable branch is depicted with red line. Point $C$ is the stability turning point of the other family of spherical orbits associated with the wormhole throat.  The observer inclination angle is $\theta_0 = 90^\circ$.}
		\label{fig:S2_Cusp1}
\end{figure}

The behaviour of the curves representing the spherical photon orbits in the vicinity of the cusp is presented in Fig. $\ref{fig:S2_Cusp1}$.  After the intersection point, denoted by $A$, the branch associated with the outer light ring reaches the point B, where the spherical orbits turn stable. This corresponds to the condition $R'' (r) =0$  at $r=r_B$. At the same point the curve $\beta =\beta(\alpha)$ is not smooth, leading to the formation of another cusp. Decreasing further the radial distance, the stable branch reaches for $r=r_0$ the curve representing the spherical photon orbits at the wormhole throat at the point $C$. The intersection point $C$ corresponds also to the point, in which the spherical orbits at the throat turn stable, i.e. the condition $R'(\alpha) = 0$ is satisfied at $\alpha = \alpha_C$.

We can further show that the stability turning point $B$ is located at a constant radial distance for any spin parameter of the wormhole in the range $a\geq a^{(1)}_{crit}\approx 0.07$. The condition $R'' (r) =0$ on the spherical orbits reduces to the equation

\begin{eqnarray}
 -\Sigma \frac{\omega''}{\omega'} + \Sigma^2 + \Sigma ' =0,
\end{eqnarray}

\noindent
where the function $\Sigma$ is defined in ($\ref{curve1}$). It has a single real solution such that $r>r_0$ equal to $r_B \approx 1.398$. Nevertheless, the location of the point $B$ in the impact parameter space depends on the spin parameter through the relations ($\ref{alpha}$)-($\ref{curve1}$) for the celestial coordinates $\alpha$ and $\beta$. The location of the stability turning point $C$ for the spherical orbits on the wormhole throat can be obtained  from the condition $R'(\alpha) = 0$. It leads to the value of the impact parameter $\alpha_C = -r_0/5a$, while the corresponding value of $\beta$ is given by

\begin{eqnarray}\label{beta_C}
\beta^2_C = \frac{r^2_0}{25}\left(9e^4 -\frac{1}{a^2}\right).
\end{eqnarray}

Decreasing the angular momentum the three characteristic points $A$, $B$ and $C$ move towards the equatorial plane. During this process the contribution of the curve associated with the spherical orbits at the throat to the shadow boundary diminishes, and the shadow becomes predominantly determined by the family of spherical orbits connected with the outer light ring. The point $B$ reaches the $\alpha$-axes at the highest value of the spin parameter. It can be determined by solving the system

\begin{eqnarray}
R''(r) = 0, \quad~~~ \beta(r) =0,
\end{eqnarray}
which gives the critical value $a = a^{(1)}_{crit}\approx 0.07 $.  It can be shown that this critical value can be obtained equivalently from the system

\begin{eqnarray}
\beta'(r) = 0, \quad~~~ \beta(r) =0.
\end{eqnarray}

\begin{figure*}[t!]
    \centering
    \begin{subfigure}[t]{0.65\textwidth}
        \includegraphics[width=\textwidth]{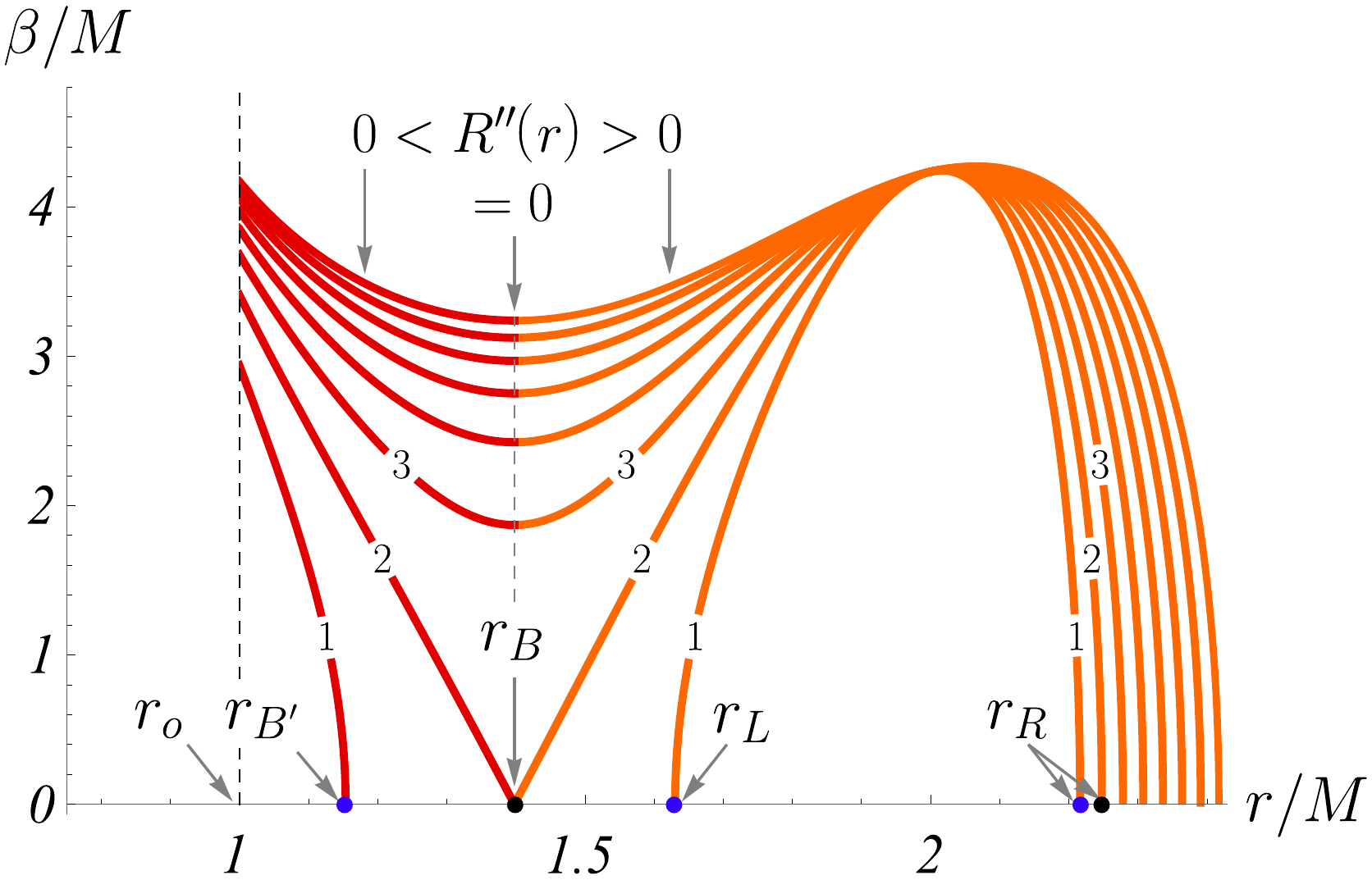}
           \end{subfigure}
           \caption{\label{fig:S2_beta}\small Dependence of the impact parameter $\beta$ on the radial distance for different spin parameters. Three representative behaviors are denoted by numbers. For spin parameters $a>a^{(1)}_{crit}\approx 0.07$ the function $\beta(r)$ has a single real root, and a minimum at $R''(r) = 0$ (curve 3). For $a=a^{(1)}_{crit}$ an additional double root arises at $R''(r) = 0$ (curve 2). For $a<a^{(1)}_{crit}$ three real roots exist corresponding to the location of unstable/stable light rings (curve 1). The unstable spherical orbits are depicted in orange, while the stable ones are in red.}
\end{figure*}

For spin parameters $a> a^{(1)}_{crit}$ the function $\beta(r)$ has a single real root $r=r_R$ in the range $r>r_0$, which determines the rightmost edge of the shadow $\alpha_R$. Furthermore, it has a minimum at the radial distance satisfying $R''(r) = 0$ (see Fig. $\ref{fig:S2_beta}$). At the critical value  $a^{(1)}_{crit}\approx 0.07$ an additional double root $r=r_B$ arises, which coincides with the minimum $R''(r) = 0$. For spin parameters $a < a^{(1)}_{crit}$ the function $\beta(r)$ possesses three real roots in the range $r>r_0$. The smallest one $r=r_{B'}$ determines the intersection point of the stable branch of the outer spherical orbits with the equatorial plane, corresponding to a stable light ring. The two larger ones $r_L$ and $r_R$ determine two unstable light rings, which correspond to the left- and rightmost edges of the shadow $\alpha_L$ and $\alpha_R$ for $a\leq0.06$.

The point $A$ reaches the equatorial plane at a second critical value $a= a^{(2)}_{crit}\approx 0.06$ obtained by the equations

\begin{eqnarray}
\beta_L(\alpha) = \beta_R(\alpha), \quad~~~ \beta_L(\alpha) =0,
\end{eqnarray}
where $\beta_L$ is the impact parameter $\beta$ on the curve associated with the spherical orbits at the throat, while $\beta_R$ is restricted to the spherical orbits connected with the outer light ring. For an equatorial observer the point $C$ coincides with the $\alpha$-axis for the lowest spin parameter $a^{(3)}_{crit} = 1/3e^2 \approx 0.0451$, which is determined by the expression ($\ref{beta_C}$) for $\beta = 0$.  These critical values of the spin parameter delimit rotation rate ranges with qualitatively different features of the spherical photon orbits.
\begin{figure}[t!]
    \centering
		\setlength{\tabcolsep}{ 0 pt }{\normalsize\tt
		\begin{tabular}{ c }
            \includegraphics[width=0.95\textwidth]{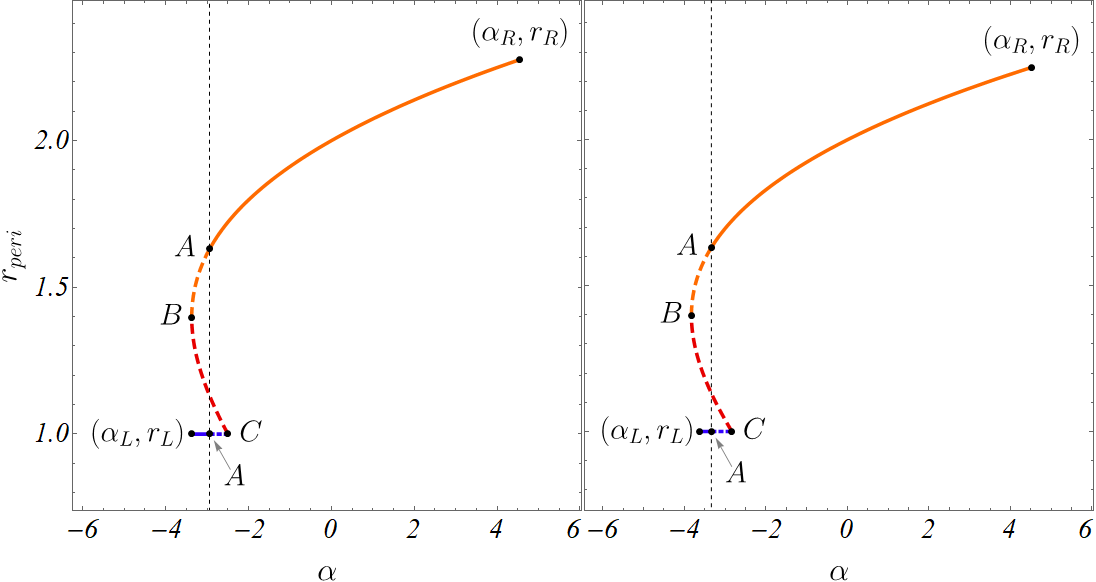}  \\[2mm]
			\hspace{1.4cm} ${a=0.08}$ \hspace{4.45cm} $a=a^{(1)}_{crit}\approx 0.07$
		\end{tabular}}
\caption{\small Dependence of the impact parameter $\alpha$ on the radial distance along the shadow boundary for spin parameter $a\geq a^{(1)}_{crit}\approx 0.07$. The observer inclination angle is $\theta_0 = 90^\circ$. The impact parameters of the left- and rightmost shadow edges are denoted by $\alpha_L$ and $\alpha_R$, respectively, while the intersection point between the two shadow boundary branches is denoted by $A$. The same color conventions are used as in Fig. $\ref{fig:S2_Cusp1}$. Points $B$ and $C$ are the stability turning points of the two families of spherical orbits. Only the unstable spherical orbits located prior to the intersection point $A$, given by solid lines, contribute to the shadow boundary (see main text).}
		\label{fig:S2_FPO}
\end{figure}

The dependence of the impact parameter $\alpha$ on the radial distance, restricted to the spherical photon orbits, is presented in Fig. $\ref{fig:S2_FPO}$. We denote by $r_{peri}$ the radial coordinate normalized by the location of the wormhole throat $r=r_0$. We observe that the radial coordinate undergoes a jump at the intersection point $A$ between the two shadow branches. As illustrated in Fig. $\ref{fig:S2_Cusp1}$, only the unstable spherical orbits located prior to the intersection point $A$ contribute to the shadow boundary. This can be explained in the following way.

\begin{figure*}[tb]
    \centering
    \begin{subfigure}[t]{0.45\textwidth}
        \includegraphics[width=\textwidth]{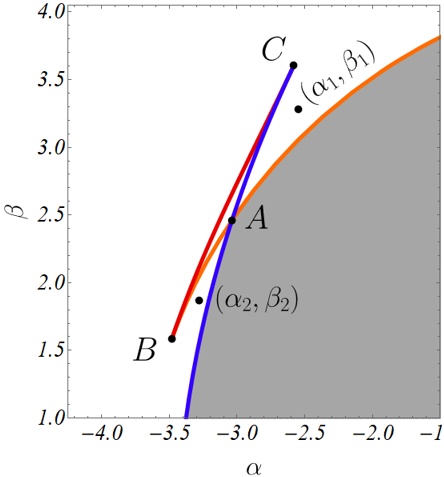}
           \end{subfigure}
           \caption{\label{fig:S2_Cusp4}\small Behavior of the spherical orbits in the vicinity of the shadow cusp for spin parameter $a > a^{(1)}_{crit}\approx 0.07$.  Geodesics with impact parameters $(\alpha_1, \beta_1)$ and $(\alpha_2, \beta_2)$ scatter away to infinity, and do not give contribution to the shadow. }
\end{figure*}

The curve connected with the outer light ring separates in the impact parameter space the geodesics, which scatter away from the corresponding maximum in the effective potential, from those that pass through it. Similarly, the other curve in the shadow boundary differentiates between geodesics which scatter away or pass through the maximum located at the wormhole throat.  Then, a geodesic with impact parameters $(\alpha_2, \beta_2)$ located as demonstrated in Fig. $\ref{fig:S2_Cusp4}$ will pass through the outer maximum but scatter away from the inner maximum at the wormhole throat, and escape to infinity. On the other hand, a geodesic with impact parameters $(\alpha_1, \beta_1)$ according to same figure will scatter away already from the outer maximum without reaching the maximum at the throat. Therefore, geodesics with such impact parameters do not contribute to the wormhole shadow. The unstable spherical orbits located in the impact parameter space after the intersection point $A$ do not influence the shadow. Yet, they have impact on the lensing properties of the spacetime, as observed in \cite{Cunha:2017b}.

\subsubsection*{2. Wormhole spin $a^{(1)}_{crit}\approx 0.07> a\geq a^{(2)}_{crit}\approx 0.06$}

After the  critical value $a = a^{(1)}_{crit}\approx 0.07$ the curve associated with the outer light ring closes, but the shadow boundary is still determined by the two families of unstable photon orbits. At the same time the stable branch of the photon orbits associated with the outer light ring gets disconnected from the unstable one. This feature can be seen in the behavior of the impact parameter $\alpha$ as a function of the radial distance illustrated in Fig. $\ref{fig:S2_FPO}$, where we observe a gap. In this regime the point $B'$ is not connected to a stability turning point, but denotes the intersection of the stable branch with the $\alpha$-axis, corresponding to a stable light ring. It coincides with the smallest root of the function $\beta (r) =0$ in the range $r > r_0$ (see Fig. $\ref{fig:S2_beta}$).

When the spin parameter decreases the stable branch shifts to the left, so that the intersection point $B'$ occurs at smaller values of $\alpha = \alpha_{B'} $. At the same time the intersection point $A$ moves towards the $\alpha$-axis reaching it at $a= a^{(2)}_{crit}\approx 0.06$ for $\alpha_A > \alpha_{B'}$.
\begin{figure}[t!]
 \centering
		\setlength{\tabcolsep}{ 0 pt }{\normalsize\tt
		\begin{tabular}{ c }
            \includegraphics[width=0.9\textwidth]{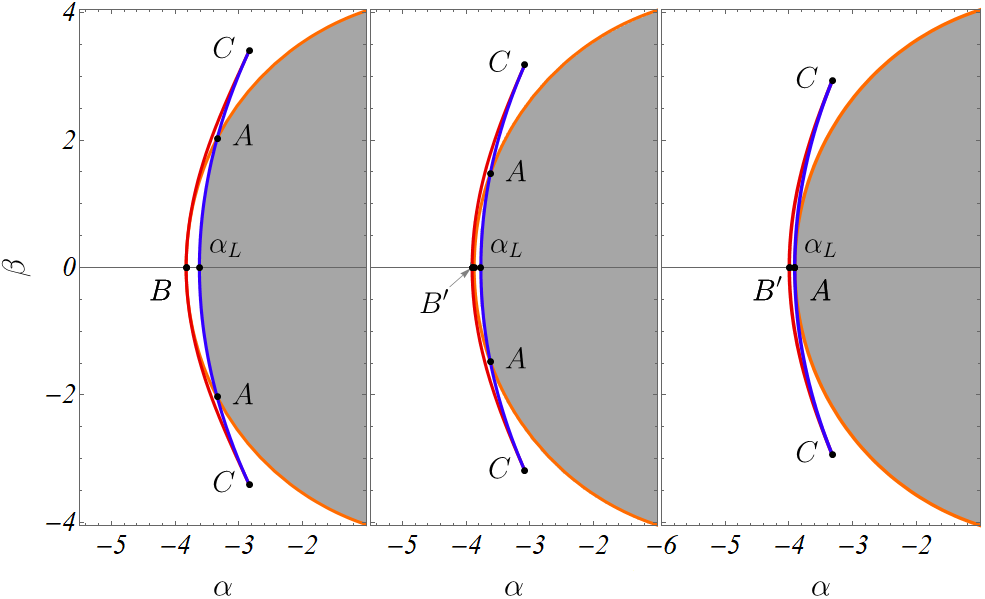} \\[1mm]
            \hspace{1.2cm}	$a=0.07$ \hspace{2cm} $a=0.065$ \hspace{1.6cm} $a=a^{(2)}_{crit}\approx0.06$
		\end{tabular}}
\vspace{-2.0mm}
\caption{\small Behavior of the spherical orbits in the vicinity of the shadow cusp for spin parameter $a^{(2)}_{crit}\approx 0.06 \leq a < a^{(1)}_{crit}\approx 0.07$.  The observer inclination angle is $\theta_0 = 90^\circ$. The stable spherical orbits associated with the outer light ring (red line) get disconnected from the unstable ones (orange line). Their intersection point with the equatorial plane, corresponding to a stable light ring, is denoted by $B'$. }
		\label{fig:S2_Cusp2}
\end{figure}
\begin{figure}[t!]
    \centering
		\setlength{\tabcolsep}{ 0 pt }{\normalsize\tt
		\begin{tabular}{ c }
            \includegraphics[width=0.87\textwidth]{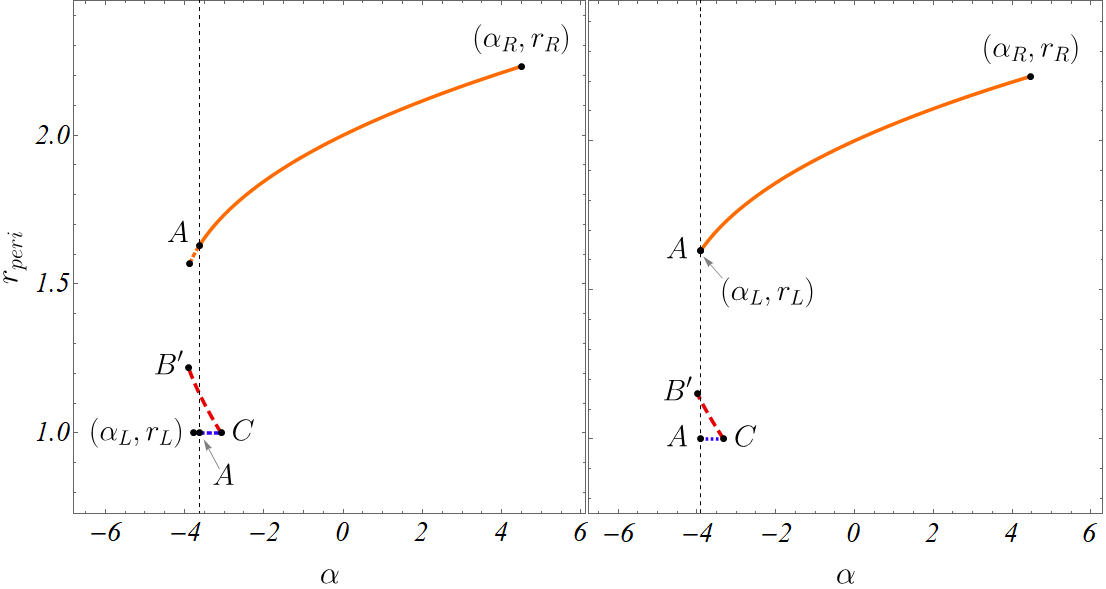}  \\[1mm]
			\hspace{0.7cm} ${a=0.065}$ \hspace{4.3cm} $a=0.06$
		\end{tabular}}
\vspace{-2.0mm}
\caption{\small Dependence of the impact parameter $\alpha$ on the radial distance at the shadow boundary for spin parameter $a^{(2)}_{crit}\approx 0.06 \leq a < a^{(1)}_{crit}\approx 0.07$. The observer inclination angle is $\theta_0 = 90^\circ$.  The same  conventions are used as in Fig. $\ref{fig:S2_Cusp2}$. The stable and unstable branches of the spherical orbits associated with the outer light ring get disconnected, and we observe a gap.}
\end{figure}

\subsubsection*{3. Wormhole spin $a < a^{(2)}_{crit}\approx 0.06$}

\begin{figure}[t!]
 \centering
		\setlength{\tabcolsep}{ 0 pt }{\normalsize\tt
		\begin{tabular}{ c }
           \includegraphics[width=0.98\textwidth]{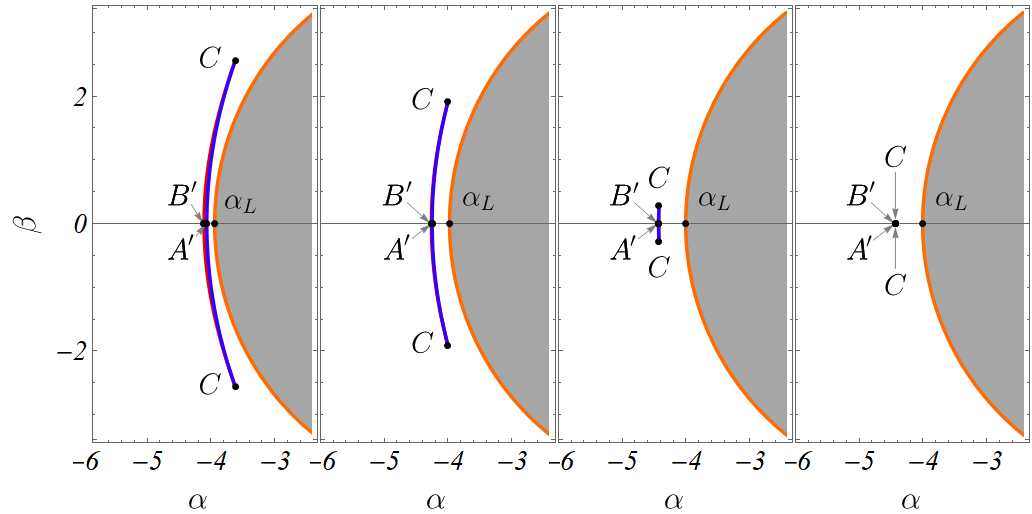} \\[2mm]
          		 \hspace{1.5cm}	$a=0.055$ \hspace{1.5cm} $a=0.05$ \hspace{1.3cm} $a=0.0452$ \hspace{0.4cm} $a=a^{(3)}_{crit}\approx0.0451$
		\end{tabular}}
\caption{\small Behavior of the spherical orbits in the vicinity of the shadow cusp for spin parameter $a < a^{(2)}_{crit}\approx 0.06$. The observer inclination angle is $\theta_0 = 90^\circ$. The curves denoting unstable spherical orbits at the throat (blue line), and the stable spherical orbits associated with the outer light ring (red line) are too close to each other to be distinguished.}
		\label{fig:S2_Cusp3}
\end{figure}
\begin{figure}[h!]
\vspace{0.75cm}
    \centering
		\setlength{\tabcolsep}{ 0 pt }{\normalsize\tt
		\begin{tabular}{ c }
            \includegraphics[width=0.9\textwidth]{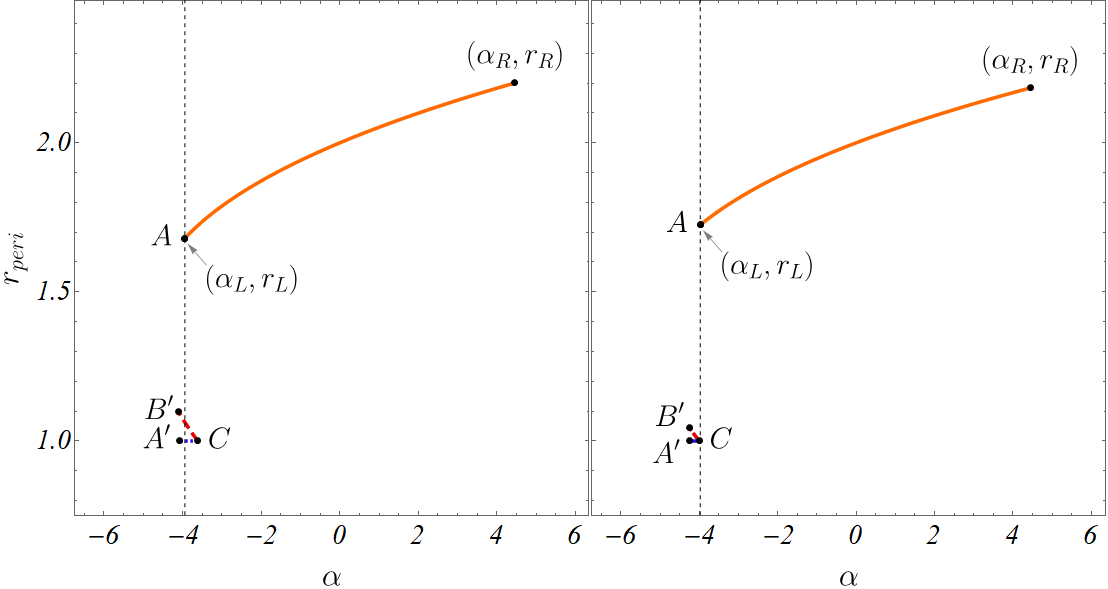}  \\[2mm]
            \hspace{0.7cm} $a=0.055$ \hspace{4.5cm} $a=0.05$
		\end{tabular}}
\caption{\small Dependence of the impact parameter $\alpha$ on the radial distance at the shadow boundary for spin parameter $a < a^{(2)}_{crit}\approx 0.06$.}
\end{figure}

At the critical value $a = a^{(2)}_{crit}\approx 0.06$ the curve determined by the unstable spherical orbits at the throat gets tangent to the curve associated with the outer light ring. They touch at the point $A$, which is now located at the equatorial plane. For values of the spin parameter lower than $a^{(2)}_{crit}$, the two families of unstable spherical orbits get disconnected, and the spherical orbits at the throat stop giving contribution to the wormhole shadow. The shadow is completely determined by the closed curve associated with the outer unstable spherical orbits. In this regime we denote by $A'$ the intersection point of the curve connected with the unstable spherical orbits at the throat with the equatorial plane.  Decreasing the angular momentum the stable branch connected with the outer spherical orbits, and the unstable branch at the throat shift to lower values of $\alpha$, however at a different rate. The point $A'$ moves faster than the point $B'$, and the two curves approach each other. At the same time they stay connected at the point $C$, and their length simultaneously reduces, while the point $C$ approaches the equatorial plane.  At the critical value of the spin parameter $a = a^{(3)}_{crit}\approx 0.0451$ the point $C$ reaches the equatorial plane, and the point $A'$ reaches the point $B'$, all three points coinciding at $\alpha_C = \alpha_{A'} = \alpha_{B'}$ (see Fig. $\ref{fig:S2_Cusp3}$). The coordinate of the leftmost edge of the shadow $\alpha_L$ also decreases approaching in the limit $a \rightarrow 0$ the location of the photon sphere of the static wormhole $\alpha^{st}_L = -2r_0e^{\frac{3}{4}}\approx - 4.234$.

\section{Conclusion}

In this paper we consider the shadow of a class of rotating traversable wormholes. The null geodesic equations are separable in the corresponding spacetime, and the shadow boundary can be obtained analytically. It is constructed as a superposition of two families of unstable spherical orbits, associated with two light rings. For some of the wormhole solutions the two shadow branches merge continuously, while for others the radial coordinate undergoes a jump at the junction, leading to the formation of cusps in the shadow edge. Such cuspy shadows were observed previously for Kerr black holes with Proca hair, and for certain phenomenological black holes in the alternative theories of gravity. In our work we provide an example for a horizonless compact object with a similar shadow structure.

We investigate the structure of the stable spherical photon orbits in the vicinity of the cusps, and the cusp dynamics with the variation of the angular momentum for an equatorial observer. For high wormhole spins stable spherical orbits exist, which are not connected with a stable light ring. Decreasing the angular momentum a critical spin is achieved, for which the stable branch crosses the equatorial plane, and a stable light ring arises. At even lower spins a second critical value is reached when the cusps disappear, and the wormhole shadow tends to circular.

\section*{Acknowledgments}
P.N. gratefully acknowledges support by the Bulgarian NSF Grant $\textsl{DM 18/3}$.


\begin{thebibliography}{tbds}


\bibitem{Bardeen}
J. M. Bardeen,
``Timelike and null geodesics in the Kerr metric,'' in {\it Black Holes}, editors Dewitt, C. and Dewitt, B. S., pp. 215-239, Gordon and Breach, New York (1973).


\bibitem{Nedkova:2013}
P. Nedkova, V. Tinchev, S. Yazadjiev,
``Shadow of a rotating traversable wormhole'',
Phys. Rev. D88 (2013) 124019.


\bibitem{Sakai}
N. Sakai, H. Saida, and T. Tamaki,
``Gravastar shadows'',
Phys. Rev. D 90 (2014) 104013.

\bibitem{Cunha:2018}
P. Cunha, C. Herdeiro, M. J. Rodriguez ,
``Does the black hole shadow probe the event horizon geometry?'',
Phys.Rev. D97 (2018) 084020.


\bibitem{Doeleman}
S. Doeleman, E. Agol, D. Backer et al., ``Imaging an event horizon: submm-VLBI of a super massive black hole'', Astro2010: The Astronomy and Astrophysics Decadal Survey, Science White Papers, No. 68 (2009).

\bibitem{Grenzebach:2014}
A. Grenzebach, V. Perlick and L\"{a}mmerzahl,
``Photon regions and shadows of Kerr-Newman-NUT black holes with a cosmological constant'',
Phys. Rev. D89 (2014) 124004.

\bibitem{Grenzebach:2015a}
A. Grenzebach,
``Aberrational effects for shadows of black holes'',
 Fund. Theor. Phys. 179 (2015) 823.

\bibitem{Grenzebach:2015b}
A. Grenzebach, V. Perlick and L\"{a}mmerzahl,
``Photon regions and shadows of accelerated black holes '',
Int. J. Mod. Phys. D24 (2015) 1542024.

\bibitem{Perlick}
V. Perlick, Oleg Yu. Tsupko, G. Bisnovatyi-Kogan,
``Influence of a plasma on the shadow of a spherically symmetric black hole'',
Phys.Rev. D92 (2015) 104031.


\bibitem{Rezzolla:2015}
 A. Abdujabbarov, L. Rezzolla, B.  Ahmedov, ``A coordinate independent characterisation of a black-hole shadow'',
 Mon. Not. Roy. Astron. Soc. 454 (2015) 2423.

\bibitem{Tinchev:2014}
V. Tinchev, S. Yazadjiev,
``Possible imprints of cosmic strings in the shadows of galactic black holes'',
Int. J. Mod. Phys. D23 (2014) 1450060.

\bibitem{Wang:2018}
M. Wang, S. Chen, J. Jing,
``Chaotic shadow of a non-Kerr rotating compact object with quadrupole mass moment'',
	arXiv:1801.02118 [gr-qc].


\bibitem{Amarilla:2010}
L. Amarilla, E. F. Eiroa and G. Giribet,
``Null geodesics and shadow of a rotating black hole in extended Chern-Simons modified gravity'',
Phys. Rev. D 81 (2010) 124045.

\bibitem{Amarilla:2012}
L. Amarilla and E. F. Eiroa,
``Shadow of a rotating braneworld black hole,''
Phys. Rev. D 85 (2012) 064019.

\bibitem{Amarilla:2013}
L. Amarilla and E. F. Eiroa
``Shadow of a Kaluza-Klein rotating dilaton black hole '',
Phys.Rev. D87 (2013)  044057.

\bibitem{Cunha:2016a}
P. V.P. Cunha, C. Herdeiro, B. Kleihaus, J. Kunz, E. Radu
``Shadows of Einstein-dilaton-Gauss-Bonnet black holes'',
Phys.Lett. B768 (2017) 373.

\bibitem{Johannsen}
T. Johannsen,
``Testing the no-hair theorem with observations of black holes in the electromagnetic spectrum'',
Class. Quant. Grav. 33 (2016) 124001.

\bibitem{Broderick}
A. E. Broderick , T. Johannsen, A. Loeb, D. Psaltis,
``Testing the no-hair theorem with event horizon telescope observations of Sagittarius A*'',
Astrophys.J. 784 (2014) 7.


\bibitem{Bambi:2015}
C. Bambi,
``Testing the Kerr paradigm with the black hole shadow'', [arXiv:1507.05257].

\bibitem{Tsukamoto}
N. Tsukamoto, Z. Li and C. Bambi,
``Constraining the spin and the deformation parameters from the black hole shadow'',
JCAP 1406 (2014) 043.

\bibitem{Takahashi}
R. Takahashi,
``Shapes and positions of black hole shadows in accretion disks and spin parameters of black holes'',
Astrophys.J. 611 (2004) 996.

\bibitem{Hioki}
K. Hioki, K. Maeda,
``Measurement of the Kerr spin parameter by observation of a compact object's shadow'',
Phys. Rev. D 80 (2009) 024042.

\bibitem{Li}
Z. Li, C. Bambi
``Measuring the Kerr spin parameter of regular black holes from their shadow'',
JCAP 1401 (2014) 041.



\bibitem{Yumoto}
A. Yumoto, D. Nitta, T. Chiba, N. Sugiyama,
``Shadows of multi-black holes: analytic exploration'',
Phys. Rev. D 86 (2012) 103001.

\bibitem{Nitta}
 D. Nitta, T. Chiba, N. Sugiyama,
``Shadows of colliding black holes'',
 Phys.Rev. D84 (2011) 063008.

\bibitem{Shipley}
J. O. Shipley, S. Dolan,
``Binary black hole shadows, chaotic scattering and the Cantor set'',
Class. Quant. Grav. 33 (2016) 175001.

\bibitem{Cunha:2015}
P. V. P. Cunha, C. A. R. Herdeiro, E. Radu, H. F. Runarsson,
``Shadows of Kerr black holes with scalar hair'',
Phys. Rev. Lett. 115 (2015) 211102.

\bibitem{Cunha:2016}
P. V. P. Cunha, J. Grover, C. Herdeiro, E. Radu, H. Runarsson, A. Wittig,
``Chaotic lensing around boson stars and Kerr black holes with scalar hair'',
Phys. Rev. D 94 (2016) 104023.

\bibitem{Bohn}
Bohn, et. al. , ``What does a binary black hole merger look like?'', Class. Quant. Grav. 32 (2015) 065002.

\bibitem{Abdolrahimi:2015b}
S. Abdolrahimi, R. Mann, C. Tzounis,
``Distorted local shadows'',
Phys. Rev. D 91 (2015) 084052.

\bibitem{Abdolrahimi:2015c}
S. Abdolrahimi, R. Mann, C. Tzounis,
``Double images from a single black hole'',
Phys. Rev. D 92 (2015) 124011.

\bibitem{Grover:2018}
J. Grover, J. Kunz, P. Nedkova, A. Wittig, S. Yazadjiev,
``Multiple shadows from distorted static black holes'',
Phys.Rev. D 97 (2018) 084024.

\bibitem{Shaikh}
R. Shaikh,
``Shadows of rotating wormholes'',
arXiv:1803.11422.


\bibitem{Cunha:2017b}
P. Cunha, C. Herdeiro, E. Radu,
``Fundamental photon orbits: black hole shadows and spacetime instabilities'',
Phys. Rev.  D 96 (2017)  024039.

\bibitem{Wang}
M. Wang, S. Chen, J. Jing,
``Shadow casted by a Konoplya-Zhidenko rotating non-Kerr black hole'',
JCAP 1710 (2017) 051.

\bibitem{Teo:1998}
E. Teo,
``Rotating traversable wormholes'',
Phys.Rev. D58 (1998) 024014.
\end{thebibliography}
\end{document}